# Intercorrelated ferroelectrics in 2D van der Waals materials

Yan Liang, Shiying Shen, Baibiao Haung, Ying Dai,* Yandong Ma*

School of Physics, State Key Laboratory of Crystal Materials, Shandong University, Shandanan Str. 27, Jinan 250100, People's Republic of China

*Corresponding author: daiy60@sina.com (Y.D.); yandong.ma@sdu.edu.cn (Y.M.)

**ABSTRACT**

2D intercorrelated ferroelectrics, exhibiting a coupled in-plane and out-of-plane ferroelectricity, is a fundamental phenomenon in the field of condensed-mater physics. The current research is based on the paradigm of bi-directional inversion asymmetry in single-layers, which restricts 2D intercorrelated ferroelectrics to extremely few systems. Herein, we propose a new scheme for achieving 2D intercorrelated ferroelectrics using van der Waals (vdW) interaction, and apply this scheme to a vast family of 2D vdW materials. Using first-principles, we demonstrate that 2D vdW multilayers—for example, BN, $MoS_2$, InSe, CdS, $PtSe_2$, $Tl_2O$, $SnS_2$, $Ti_2CO_2$ *etc.*—can exhibit coupled in-plane and out-of-plane ferroelectricity, thus yielding 2D intercorrelated ferroelectric physics. We further predict that such intercorrelated ferroelectrics could demonstrate many distinct properties, for example, electrical full control of spin textures in trilayer $PtSe_2$ and electrical permanent control of valley-contrasting physics in four-layer $VS_2$. Our finding opens a new direction for 2D intercorrelated ferroelectric research.

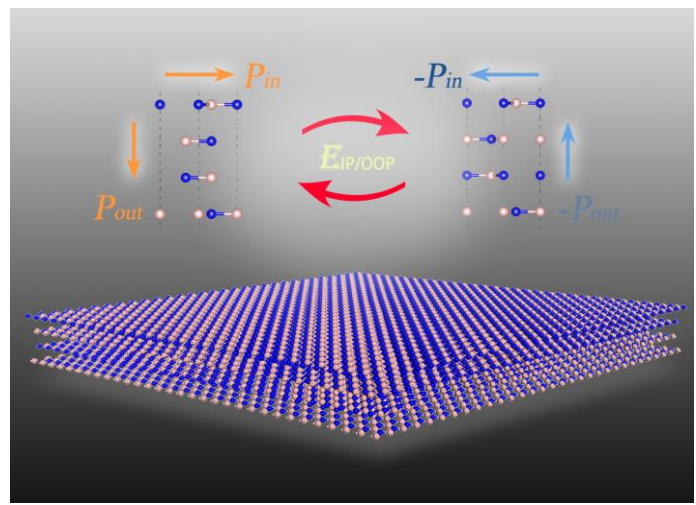

**INTRODUCTION**

The recent discovery of two-dimensional (2D) intercorrelated ferroelectrics has attracted tremendous interest in condensed matter physics [1-5]. The defining signature of 2D intercorrelated ferroelectrics is the coexistence of spontaneous in-plane (IP) and out-of-plane (OOP) polarizations, which are strongly coupled with each other. Especially, distinct from conventional ferroelectrics, the flipping of IP electric polarization can be achieved by an OOP bias, or vice versa. This characteristic feature suggests new device paradigms beyond conventional ferroelectrics—including co-planar ferroelectric field-effect transistor, multiple resistance states switching and ferroelectric rectifier [6-9]—thus holding great potential for applications in advanced information storage [1,5,10,11]. Though highly valuable, to realize 2D intercorrelated ferroelectrics, it requires not only the coexistence of IP and OOP polarizations, but also the same polarization mechanism for them, i.e., IP (OOP) dipoles being of strongly locked by OOP (IP) symmetry. Obviously, these requirements are extraordinarily stringent to be satisfied in 2D limit, as 2D IP and OOP ferroelectrics are rare themselves. So far, the existence of 2D intercorrelated ferroelectrics is only observed in single-layer α-$In_2Se_3$ [1,5,8,10,12], which greatly obstructs further experimental studies and possible applications.

Recently, numerous approaches are reported for realizing ferroelectricity, such as doping [13] functionalization [14], strain [15] vacancy [16,17], foreign-atom decoration [18,19] and edges [20]. These strategies indeed succeed in producing IP and OOP ferroelectric polarizations, but not simultaneously, and even not to say their coupling, thereby being of useless for realizing 2D intercorrelated ferroelectrics. Moreover, these approaches are difficult to perform experimentally, and are far from controllable in practice. As an alternative, superlattices, such as $(BaTiO_3)_m$-$(SrTiO_3)_n$, are proposed in the context of generating both IP and OOP ferroelectricity [21,22]. Nonetheless, the weak IP polarization in such superlattices is not only vulnerable to periodicity and external strain, but also independent from OOP polarization, making it not applicable for 2D intercorrelated ferroelectrics as well. To the best of our knowledge, it is still unclear how to expand the scope for 2D intercorrelated ferroelectric materials beyond α-$In_2Se_3$. It is therefore of fundamental interest to explore a new scheme for achieving 2D intercorrelated ferroelectrics.

In this work, we present a new scheme for achieving 2D intercorrelated ferroelectrics on the basis of first-principles calculations. Instead of utilizing the paradigm of non-centrosymmetric single-layers, the design principle is to use particular vdW interaction in 2D limit. By tuning the unique stacking configuration, both IP and OOP ferroelectric polarizations, as well as their strong coupling, can be obtained in 2D vdW multilayers, thus establishing 2D intercorrelated ferroelectrics. This unique scheme is successfully exemplified in a vast family of 2D vdW materials, such as BN, $MoS_2$, InSe, CdS, $PtSe_2$, $Ti_2CO_2$ *etc.*. As a new 2D intercorrelated ferroelectric family, its potential coupling with other physics may provide novel physics. We also predict that intriguing phenomena like electrical full control of spin textures and electrical permanent control of valley-contrasting physics could occur in such systems. Our results greatly extend the scope for candidate materials of 2D intercorrelated ferroelectrics and hence will simulate immediate experimental interest.

**RESULTS**

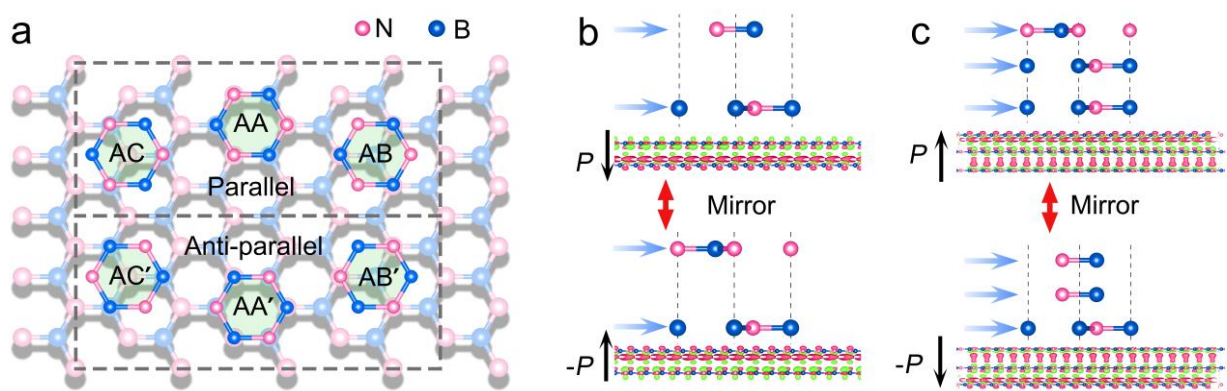

**Figure 1**. (a) Parallel/anti-parellel stacking configurations for sequential h-BN layer with respect to bottom layer. Ferroelectric switching of (b) BL h-BN and (c) TL h-BN under vertical electric field; the corresponding interlayer charge transfers are also displayed wherein red and green isosurfaces represent charge depletion and accumulation, respectively. Black and blue arrows represent the corresponding OOP polarization and i-IP directions, respectively.

Electrical polarization in 2D limit stems from the separation of positive and negative charged ions. Once the separation is switchable, 2D ferroelectricity will be realized [23,24]. According to this mechanism, inversion symmetry should be broken first. Then the mirror symmetry should be absent in the OOP and assigned IP directions, which respectively lead to the electric dipoles. To further realize the 2D intercorrelated ferroelectrics, the feasible reversal operations on the OOP and assigned IP electric dipoles should originate from identical physics. And apparently, in 2D intercorrelated ferroelectrics, the ferroelectric reversal operation should act as a 180 ° rotation with respect to the direction perpendicular to both IP and OOP polarizations. Single-layer lattice that fulfills these prerequires is extremely rare [1,25]. In this regard, going beyond the paradigm of the intrinsic particular crystal structure in single-layer lattice, new flexible alternative freedoms should be introduced to satisfy these requirements.

One such example is the vdW interaction is 2D lattice, wherein both the layer-thickness and layer-stacking can act as new freedoms to engineer the crystal symmetry and in turn the electric dipoles. Here, without loss of generality, we start with h-BN having non-ferroelectrics to dig into the design principle for realizing 2D intercorrelated ferroelectrics using vdW interactions. Due to the three-fold rotation symmetry of $D_{3h}$ point group, single-layer h-BN possesses three equivalent IP polarizations along the [110], [$\bar{2}$10], and [1$\bar{2}$0] directions. For convenience of discussion, we refer to such intrinsic IP polarization within single-layer h-BN as i-IP polarization. And the sum of these three i-IP polarizations results in a zero IP net polarization. Although substrate proximity effect and device geometry can break the three-fold rotation, yielding an IP polarization [26], the polarization is not reversible. This, combined with the mirror symmetry of $M_z$, render that 2D intercorrelated ferroelectrics could not occur in single-layer h-BN.

Before studying 2D vdW multilayers, we first define six typical stacking patterns between two layers by distinguishing atomic positions of sequential layer with respect to bottom layer, i.e., AA, AB, AC, AA′, AB′ and AC′ (**Figure 1a**). In bilayer (BL) h-BN, the stacking configurations can be divided into two groups: the "parallel" (AA, AB and AC, in which assigned i-IP polarizations of the two layers are aligned parallel) and "anti-parallel" (AA′, AB′ and AC′, in which assigned i-IP polarizations of the two layers are anti-parallel) configurations [27]. Within each group, the stacking configurations can be transformed into each other under layer sliding. For the three anti-parallel configurations, they exhibit a spatial inversion symmetry, thus excluding their possibility for either IP or OOP ferroelectricity. While for the three parallel configurations, the IP polarization arises from two parts: the assigned i-IP polarizations in each layer and the IP polarization induced by the charge redistribution between two layers (c-IP). As the assigned i-IP polarizations of the two layers are aligned parallel, the reversal of the assigned i-IP polarization through layer sliding is forbidden, thereby excluding the IP ferroelectricity. However, arising from the simultaneous absence of mirror and inversion symmetries, the AB and AC configurations present a spontaneous OOP polarization of 0.68 μC·cm$^{-2}$ and -0.68 μC·cm$^{-2}$, respectively (**Figure 1b**) [27,28]. Given the fact that these two configurations can be switched to each other via layer sliding, these two configurations correspond to two ferroelectric states, thus holding potential for OOP ferroelectricity. Upon introducing an extra layer to BL h-BN, the OOP ferroelectricity can also be obtained. For instance, through layer sliding, the AAB′ configuration with an OOP polarization pointing upwards can be transformed into the ABB configuration with an OOP polarization pointing downwards (**Figure 1c**). However, similar to the case of parallel BL h-BN where the i-IP polarization is not switchable under layer sliding, the reversal of these two trilayer (TL) configurations is equivalent to a mirror operation with respect to the middle plane, rather than a 180 ° rotation with respect to the direction perpendicular to the OOP polarization, forbidding the IP as well as intercorrelated ferroelectricity.

Accordingly, by introducing the vdW interaction, the OOP ferroelectricity that is absent in single-layer lattice can be successfully achieved in multilayer h-BN. But the existence of assigned i-IP polarization within single layers blocks the formation of IP ferroelectricity and hence intercorrelated ferroelectricity in 2D vdW multilayers. For realizing IP ferroelectricity, the assigned i-IP polarization should be offset when forming the 2D vdW multilayers, and the c-IP polarization should be preserved. Only in this case, the equivalent 180 ° rotation with respect to the direction perpendicular to both IP and OOP polarizations is possible to be obtained under layer sliding triggered by electric field. To satisfy this condition, we narrow down our choices to multilayer h-BN with even layers, and successfully realize the intercorrelated ferroelectricity in four-layer (FL) h-BN. In four-layer (FL) h-BN with intercorrelated ferroelectricity, the assigned i-IP polarizations in both two middle and two outmost layers should be "anti-parallel" aligned to offset the i-IP polarization as well as to facilitate the equivalent 180 ° rotation under layer sliding, and the inversion symmetry and mirror symmetries along OOP and assigned IP directions should be broken to produce IP and OOP polarizations. Following these two prerequisites, we consider all the possible configurations for FL h-BN and screen out 18 non-equivalent configurations with intercorrelated ferroelectricity (**Figure 2a**).

Taking the ABB′A′ configuration as an example, we explore more physics underlying the intercorrelated ferroelectricity of FL h-BN. As shown in **Figure 2b**, the central bilayer in ABB′A′ configuration exhibit an inversion symmetry. The N1 atom of layer-1 is right over the hole of the central bilayer, while the B2 atom of layer-4 locates directly below the hole. Such configuration would lead to the separation of positive and negative charge centers along the OOP direction, generating an OOP polarization pointing downwards. This is confirmed by the planar average electrostatic potentials along OOP direction shown in **Figure 2c**, wherein a discontinuity (ΔV) of 0.43 eV is formed between the vacuum levels of the top and bottom layers. The OOP polarization of the ABB′A′ configuration is found to be $3.5 \times 10^{12}$ e/cm$^2$ (0.57μC·cm$^{-2}$). Under a relative layer sliding of the central bilayer, the B1 atom of layer-1 is shifted to above the hole of the central bilayer, while the N2 atom of layer-4 is shifted to below the hole. Therefore, the relative positions of the positive and negative charge centers along the OOP direction are reversed, producing an OOP polarization of 0.57μC·cm$^{-2}$ pointing upwards. These two states can be considered as two ferroelectric states FE and FE′. To confirm the feasibility of the OOP ferroelectricity, we calculate the energy profile for the switching between them using the NEB method. As shown in **Figure 2b**, their switching is through a centrosymmetric state PE. The calculated energy barrier is 15.6 meV, comparable to that of phosphorene analogues [29] and α-In$_2$Se$_3$ [30], suggesting the feasibility of OOP ferroelectricity in ABB′A′ configuration.

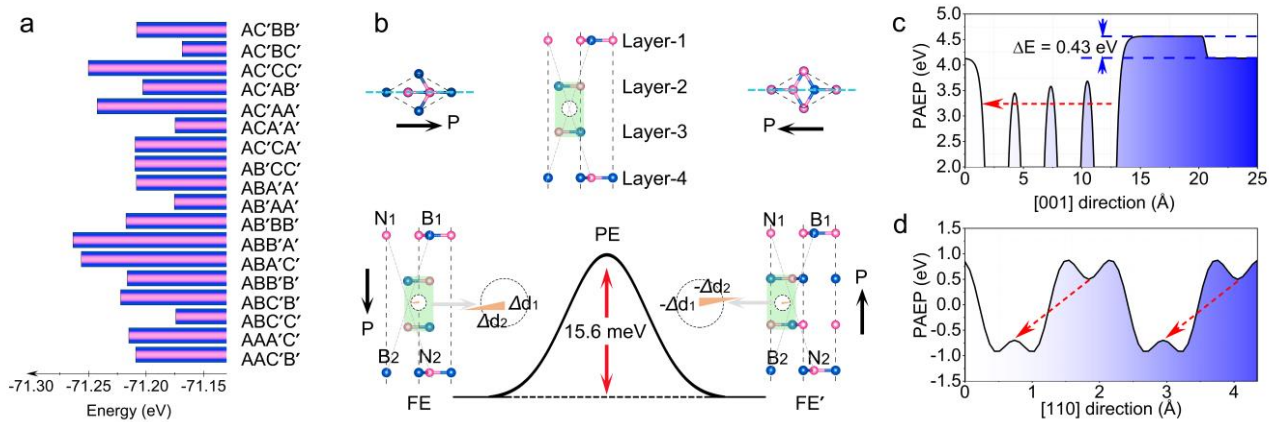

**Figure 2**. (a) Total energies per unit_cell of 18 non-equivalent configurations of FL h-BN with intercorrelated ferroelectricity. (b) Minimum energy path for polarization reversal, along with the corresponding structures, of ABB′A′ configuration of FL h-BN. Planar average electrostatic potentials of ABB′A′ configuration of FL h-BN along (c) [001] and (d) [110] directions.

Meanwhile, by focusing on the atoms on (1$\bar{1}$0) plane of the ABB′A′ configuration, as shown in **Figure 2b**, the center of the four-atom ring of central bilayer is closer to B1 atoms of layer-1 than that of B2 atom of layer-4 in the IP direction. Considering the fact that i-IP polarizations within ABB′A′ configuration are offset, such feature gives rise to the separation of positive and negative charge centers along [110] direction, yielding an IP polarization of $1.9 \times 10^{10}$ e/cm$^2$ ($3.169 \times 10^{-3}$ μC·cm$^{-2}$). This is in consistent with the unsymmetrical planar average electrostatic potential along [110] direction (**Figure 2d**). Interestingly, under the same relative layer sliding of the central bilayer, the center of the four-atom ring of central bilayer shifts closer to B2 atom of layer-4, which reverses the direction of the IP polarization to [$\bar{1}\bar{1}0$] direction (**Figure 2b**). Obviously, the energy profile for IP and OOP ferroelectric reversal should be identical, namely, the energy barrier for switching IP polarization is 15.6 meV. This suggests that the switching of IP and OOP polarizations are strongly coupled. Akin to single-layer α-In$_2$Se$_3$, there are three equivalent IP polarizations along the [110], [$\bar{2}$10], and [1$\bar{2}$0] directions, resulting in a zero IP net polarization. By utilizing the substrate proximity effect or device geometry, the three-fold IP rotation will be broken, leading to the IP ferroelectricity, which has been firmly confirmed in experiment [10,31,32]. In this case, the intercorrelated ferroelectricity is achieved in the ABB′A′ configuration of FL h-BN, i.e., the flipping of OOP/IP electric polarization can be achieved via both OOP and IP bias.

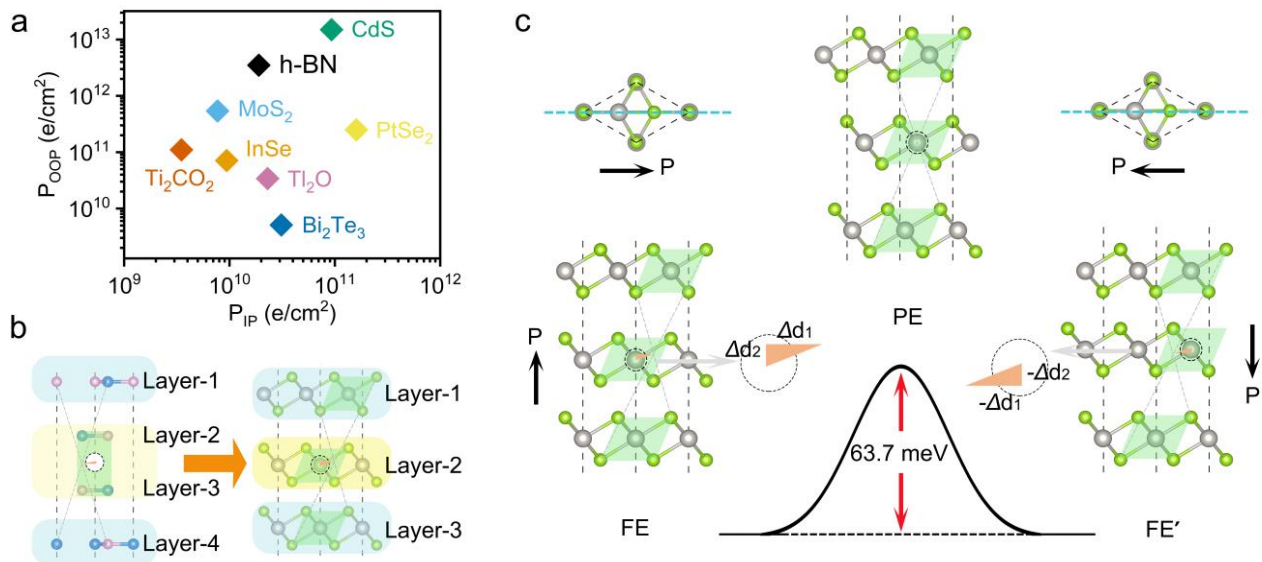

**Figure 3**. (a) IP and OOP polarizations of FL h-BN, FL MoS$_2$, FL InSe, FL CdS, TL Tl$_2$O, TL Bi$_2$Te$_3$ and TL PtSe$_2$. (b) Shematic diagram of intercorrelated ferroelectrics using three identical single-layer materials with inversion symmetry. (c) Minimum energy path for polarization reversal, along with the corresponding structures, of AAB configuration of TL PtSe$_2$.

From above, using vdW interaction, we successfully establish a design principle for realizing intercorrelated ferroelectrics in 2D lattice. It is worth emphasizing that the application of this design principle is not specific to single-layer h-BN, but essentially extendable to all single-layer materials with $D_{3h}$ symmetry, such as the transition metal dichalcogenides, group II oxides, group III-V/VI single layers and so on. Here, we take H-MoS$_2$, InSe and CdS as examples, and show their FL structures under ABB′A′ configuration in **Figure S1**. Obviously, they all harbor OOP and IP polarizations, as shown in **Figure 3a** and **Table S1**. With the relative layer sliding under electric field stimuli, just like the case of FL h-BN, the directions of both OOP and IP polarizations are reversed, acting as a 180 ° rotation with respect to the direction perpendicular to both IP and OOP polarizations. This confirms their potential for 2D intercorrelated ferroelectrics.

Actually, this scheme is also applicable for other single-layer materials with inversion symmetry, such as PtSe$_2$, Tl$_2$O, SnS$_2$, Sb$_2$Te$_2$X, MXene Ti$_2$CO$_2$, and so on. By looking at the ABB′A′ configuration of FL h-BN (**Figure 3b**), the central bilayer exhibits an inversion symmetry. And the relative layer sliding of the central bilayer will not break it. While for layer-1 and layer-4, they are required to be the same under a 180 ° rotation with respect to the direction perpendicular to both IP and OOP polarizations. These conditions can be alternatively satisfied by replacing layer-1, central bilayer and layer-4 with three identical single-layer materials with inversion symmetry. Once OOP and IP mirror symmetries are broken by tuning the stacking pattern, 2D intercorrelated ferroelectrics can also be realized in these systems. Taking TL PtSe$_2$ as an example, we illustrate the underlying physics for the reversal of OOP and IP polarizations. As shown in **Figure 3c**, the center of the four-atom ring (CF) of central layer is closer to the CF of layer-1 than that of layer-3 in both OOP and IP directions, which results in a polarization of 2.5 $\times 10^{11}$ and 1.6 $\times 10^{11}$ e/cm$^2$, respectively. Through a relative layer sliding of the central layer, both OOP and IP polarizations are reversed to the opposite directions (**Figure 3c**). The energy barrier for such switching is estimated to be 63.7 meV/f.u., suggesting the potential for 2D intercorrelated ferroelectrics. Obviously, this scenario is appliable for other single-layer materials with inversion symmetry.

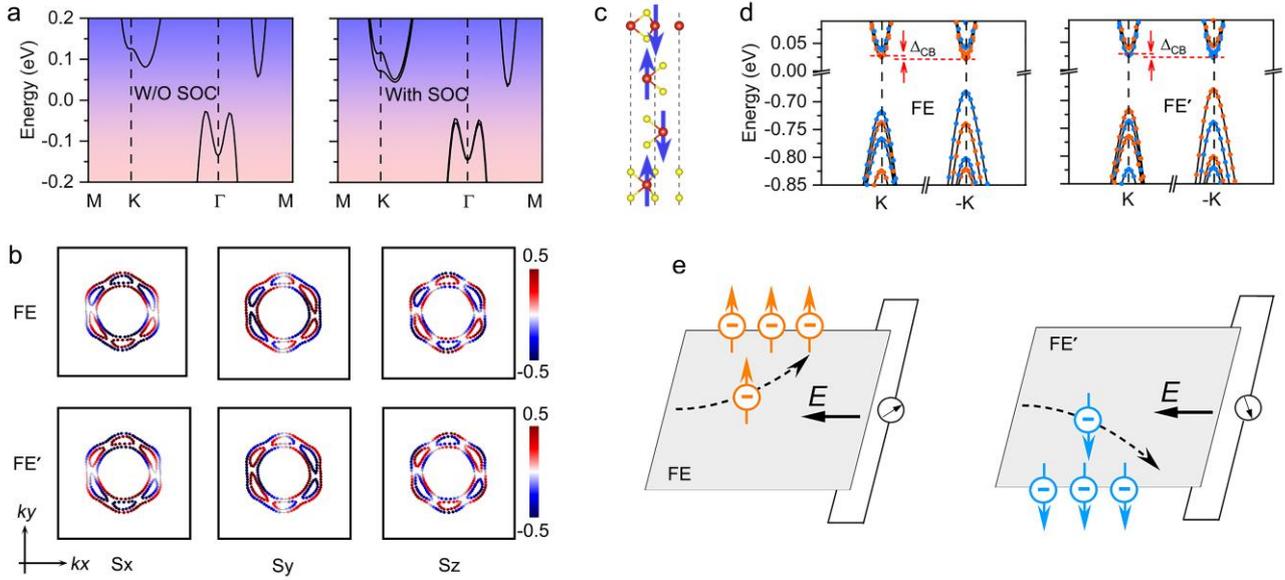

**Figure 4**. (a) Band structures of TL PtSe$_2$ without and with SOC. (b) Spin textures of valence bands around the Γ point for FE and FE′ states of TL PtSe$_2$. (c) FL VS$_2$ with interlayer AFM coupling at FE state; the blue arrows represent the spin direction of electrons on V atoms. (d) Band structures of FL VS$_2$ at FE and FE′ states with SOC; electronic states with up/down spin are represented by orange/blue colors. (e) Schematic depictions of the anomalous valley Hall effects at FE and FE′ states under electron doping and longitudinal in-plane electric field.

In principle, by utilizing vdW interaction, 2D intercorrelated ferroelectric systems can be constructed by various kinds of single-layer materials with $D_{3h}$ or inversion symmetry. This, combined with the rich and varied properties of single-layer materials, would facilitate possibility of the coupling between intercorrelated ferroelectrics and other properties, thus inspiring exciting opportunities to design novel devices with new physics and phenomenon. In the following, we respectively take Rashba effect and valley-contrasting physics as examples to address the new physics raised from their coupling with intercorrelated ferroelectrics.

Rashba effect is generally related to the presence of a potential asymmetry and a polar axis, which lifts the electron-spin degeneracy under spin-orbit coupling (SOC). In this case, significant Rashba effect is expected to be observed in 2D intercorrelated ferroelectric materials comprising heavy atoms. Different from conventional oxide films whose polarization usually is suppressed by complex depolarization field, 2D intercorrelated ferroelectric materials with robust polarization present an ideal platform for exploring the coupling between Rashba effect and intercorrelated ferroelectrics. One specific system is TL PtSe$_2$. **Figure 4a** shows the band structure of TL PtSe$_2$ without and with SOC. Upon taking SOC into account, the spin degeneracy for the highest valence band around the Γ point is lifted, producing a significant Rashba splitting $E_S$ = 11 meV. In **Figure 4b**, the upper and lower panels display the spin textures of valence bands around the Γ point for TL PtSe$_2$ with opposite ferroelectric OOP/IP polarizations. It can be seen that, when switching the OOP/IP polarization directions by either IP or OOP bias, both OOP and IP spin components of the inner and outer bands are rotated in opposite directions. This can be rationalized by the fact that the spin precession of electrons experience an effective magnetic field B = ($\boldsymbol{p} \times \boldsymbol{E}$ /2$mc^2$) [33], and OOP and IP spin components is related to the IP and OOP polar, respectively. Therefore, spin textures in TL PtSe$_2$ can be fully controlled under the action of either OOP or IP bias, which is superior to conventional 2D ferroelectrics wherein only IP or OOP spin textures reversal could occur [33,34]. Such simultaneously full control of spin textures is significance for novel spintronic applications [35].

Compared with Rashba effect, valley-contrasting physics is a more recent development. 2D hexagonal lattices with an inversion asymmetry provide an appropriate platform for operating valley-contrasting physics when their band edges are located at the corners of the Brillouin zone [36]. Especially for 2D VS$_2$ [37], LaBr$_2$ [38] and Nb$_3$X$_8$ [39], due to the strong SOC and time-reversal symmetry breaking, the desired valley polarization occurs spontaneously. When stacking these materials to form 2D intercorrelated ferroelectric systems, the coupling between valley-contrasting physics and intercorrelated ferroelectrics is expected. Here, we take ABB′A′ configuration of FL VS$_2$ as an example. In FL VS$_2$, the four component layers are anti-ferromagnetically coupled (**Figure 4c**), but the magnetic moments are inequivalent due to electric polarization. This results in a net magnetic moment of 0.004 μ$_B$, suggesting a ferrimagnetic ground state with upwards magnetization. **Figure S2** shows the band structure of FL VS$_2$ without SOC. Its CBM locates at the K/-K point, forming two valleys. When including SOC, the -K valley in the spin-up band lies lower in energy than the K valley, producing a spontaneous valley polarization (Δ$_{CB}$ = $E_K$ – $E_{-K}$) of 3.76 meV (**Figure 4d**). This value is larger than that of the experimentally demonstrated WSe$_2$/CrI$_3$ heterostructure [40,41]. In presence of a longitudinal in-plane electric field, Bloch electrons in the -K valleys will acquire an anomalous velocity ($\boldsymbol{v}_a \sim \boldsymbol{E} \times \Omega_{(\boldsymbol{k})}$) and accumulate on the left side of the sample when shifting the Fermi level between the K and -K valleys via chemical doping or electrical gating (**Figure 4e**).

After switching the OOP/IP polarization directions by either IP or OOP bias, the magnetization of FL VS$_2$ shifts downwards. And the spin component of the K and -K valleys is reversed to spin-down; however, the -K valley from the spin-down band is still lower in energy than the K valley, preserving the valley polarization of 3.76 meV (**Figure 4d**). While for Berry curvature, as shown in **Figure S3**, distinct from case with magnetic field [38], the absolute values remain unchanged, but the sign is reversed, at the K and -K valleys under the OOP/IP polarization switching. As a result, in presence of a longitudinal in-plane electric field, Bloch electrons in the -K valley of the FE′ state will accumulate on the right side of the sample when shifting the Fermi level between the K and -K valleys (**Figure 4e**). Accordingly, the valley-contrasting physics, as well as the ferrimagnetism, can be well controlled in a feasible way in FL VS$_2$. We wish to stress that, although many approaches have been proposed to control the valley-contrasting physics, including circularly polarized light [42], magnetic/electric field [43,44], and substrate effect [45], most of them suffer from low efficiency, high power consumption, increased devices size and volatility. The coupling between intercorrelated ferroelectrics and valley-contrasting physics can circumvent these challenges and is of remarkably importance for developing 2D vallleytronic physics and devices.

## CONCLUSION

To summarize, a new scheme for realizing 2D intercorrelated ferroelectrics is proposed using first-principles. Going beyond the paradigm of non-centrosymmetric single-layers, our design principle is based on the particular vdW interaction in 2D limit. Both IP and OOP ferroelectric polarizations, as well as their strong coupling, can be obtained in 2D vdW multilayers, through tuning the layer number and stacking configurations, which establishes the long-pursuit 2D intercorrelated ferroelectrics. We reveal that this unique scheme can be exemplified in a vast family of 2D vdW materials, such as BN, MoS$_2$, ZnO, InSe, SiC, CdS, PtSe$_2$, TI$_2$O, SnS$_2$, Ti$_2$CO$_2$ *etc.*. More importantly, we find that as a new 2D intercorrelated ferroelectric family, they could demonstrate many unconventional phenomena and possibilities for practical applications, for example, electrical full control of spin textures in TL PtSe$_2$ and electrical permanent control of valley-contrasting physics in FL VS$_2$. Our findings greatly extend the scope for candidate materials of 2D intercorrelated ferroelectrics and provide a novel platform towards exploring novel physics.

## ACKNOWLEDGEMENT

This work is supported by the National Natural Science Foundation of China (No. 11804190), Shandong Provincial Natural Science Foundation of China (Nos. ZR2019QA011 and ZR2019MEM013), Shandong Provincial Key Research and Development Program (Major Scientific and Technological Innovation Project) (No. 2019JZZY010302), Shandong Provincial Key Research and Development Program (No. 2019RKE27004), Qilu Young Scholar Program of Shandong University, and Taishan Scholar Program of Shandong Province.